\RequirePackage{fix-cm}
\documentclass[smallextended]{svjour3}       
\smartqed  

\usepackage{latexsym}
\usepackage{amsmath}
\usepackage{amssymb}
\usepackage{ulem}
\usepackage{fancyhdr}
\usepackage{url}
\usepackage{grffile}
\usepackage[vcentering,dvips]{geometry}
\usepackage{sidecap}

\usepackage{graphicx}
\begin{document}

\title{Effect of a low density dust shell on the propagation of gravitational waves
}


\author{Nigel~T. Bishop$^*$   \and
        Petrus~J. van der Walt \and
        Monos Naidoo.
}


\institute{Nigel~T. Bishop$^*$ \at
              Department of Mathematics, Rhodes University, Grahamstown, 6140, South Africa \\
              \email{n.bishop@ru.ac.za}           
             \and
			Petrus~J. van der Walt \at
			Department of Mathematics, Rhodes University, Grahamstown, 6140, South Africa \\
			\email{p.vanderwalt@ru.ac.za}          
            \and
             Monos Naidoo \at
             Department of Mathematics, Rhodes University, Grahamstown, 6140, South Africa \\
              \email{monos.naidoo@gmail.com}           
}

\date{Received: date / Accepted: date}

\maketitle

\begin{abstract}
Using the Bondi-Sachs formalism, the problem of a gravitational wave source surrounded by a spherical dust shell is considered. Using linearized perturbation theory, the geometry is found in the regions: in the shell, exterior to the shell, and interior to the shell. It is found that the dust shell causes the gravitational wave to be modified both in magnitude and phase, but without any energy being transferred to or from the dust.
\end{abstract}

\section{Introduction}
\label{intro}

Calculations of gravitational waves (GWs), both analytical and numerical, normally assume that they propagate from source to a detector on Earth in a vacuum spacetime. Although the average cosmological density of baryonic plus dark matter is small, of order $10^{-29}$g/cm${}^3$, a detected GW event may be a considerable distance away from its source, up to order $1$ Gpc, and the quantity of intervening matter is not negligible. Further, it is possible that the astrophysical environment of a source event could be such that the source is surrounded by a substantial amount of matter. Thus, as we move into an era of precision GW measurements, it is important to quantify any effects due to propagation of GWs through a non-vacuum spacetime.

These issues have been investigated previously. There is a simple physical argument that an ideal fluid should not extract energy from a GW, because there is no physical mechanism for it to do so; and this idea has been given a precise expression in the work of Esposito~\cite{Esposito71b}, and of Ehlers {\it et al.}~\cite{Ehlers87,Ehlers96}. However, if the matter is dissipative, e.g. through shear viscosity, then one would expect GWs to be attenuated. Hawking~\cite{Hawking66} investigated  GWs in cosmological models and determined conditions for complete absorption; subsequently, the general theory of GW propagation through a viscous fluid was further developed~\cite{Esposito71a,Madore73,Anile78,Prasanna99}. More recently, Goswami {\it et al.}~\cite{Goswami17} have investigated whether the properties of dark matter can be constrained by the attenuation effect and GW observations. Baym {\it et al.}~\cite{Baym17} have extended the hydrodynamics model of the cosmological fluid to a kinetic model with low collision rates and calculated the attenuation effect.

This work uses the Bondi-Sachs~\cite{Bondi62,Sachs62} formalism for the Einstein equations; see also the reviews~\cite{Winicour05,Bishop2016a}.
We consider small perturbations about a fixed background, which topic was first considered some time ago and known as the ``quasi-Netwonian'' or ``quasi-spherical'' approximation\cite{Winicour83,Bishop96}; there has also been previous work using this approach in which a dust cloud source was considered~~\cite{Isaacson85}. However,
this work uses the method of separation of variables to construct, within a Bondi-Sachs framework, eigensolutions for linearized perturbations~\cite{Bishop-2005b}; when the background is Minkowski spacetime, these solutions have a remarkably simple analytic form. This approach has given additional insights in another context, that of GWs propagating in de Sitter spacetime~\cite{Bishop2016}; see also~\cite{Ashtekar-2015-2,Ashtekar-2015-3}.

The paper considers the model problem of a GW source in a spacetime that is empty apart from matter contained in a thin shell around the source (then, results for a thick matter shell can be modelled by adding up, i.e. integrating, over thin shells). As a first step, the shell is given the simplest equation of state, i.e., that of dust. It is found that the effect of the shell is to modify the outgoing GWs in both phase and magnitude, although in a way that does not contradict previous results about energy transfer. The modification of the GW is small, and in a cosmological context is not expected to be measurable; but it is possible that a GW event could occur in which the local astrophysical environment is such that the effect would be measurable. Further, there is a view that LIGO data for black hole mergers may contain echoes, and explanations investigated, using numerical simulations, have included new physics near the event horizon, and the astrophysical environment such as a shell around the system; see, e.g.,~\cite{Conklin2017,Konoplya2019}; this matter is discussed further in Sec.~\ref{s-conc}. In any case, the results are certainly of interest to the theory of GWs propagating in matter.

Sec.~\ref{s-back} specifies the problem, and constructs the background solution, i.e. when the geometry is spherically symmetric. The solution when the source is emitting GWs is then constructed in Sec.~\ref{s-pert}. The physical interpretation of the modified GWs is discussed in Sec.~\ref{s-phys}. The main part of the paper ends with a Conclusion (Sec.~\ref{s-conc}), which also includes a discussion of further work. The paper makes substantial use of computer algebra; the scripts used are available online, and are summarized in Appendix~\ref{a-compalg}.

\section{Background solution}
\label{s-back}
The Bondi-Sachs formalism for the Einstein equations is well-known~\cite{Bondi62,Sachs62}. The coordinates are based on outgoing null hypersurfaces labelled by the coordinate $x^0=u$. Let $x^A$ ($A=2,3$) be angular coordinates (e.g. spherical polars $(\theta,\phi)$) that label the null ray generators of a hypersurface $u=$ constant, and let $x^1=r$ be a surface area radial coordinate. The Bondi-Sachs metric describes a general spacetime, and here we write it as

\begin{align}
ds^2  = & -\left(e^{2\beta}\left(1 + \frac{W}{ r}\right)
	- r^2h_{AB}U^AU^B\right)du^2
	- 2e^{2\beta}dudr \nonumber \\
	& - 2r^2 h_{AB}U^Bdudx^A
	+  r^2h_{AB}dx^Adx^B\,,
\label{eq:bmet}
\end{align}

where $h^{AB}h_{AB}=\delta^A_C$, and the condition that $r$ is a surface area coordinate implies $\det(h_{AB})=\det(q_{AB})$ where $q_{AB}$ is a unit sphere metric (e.g. $d\theta^2+\sin^2\theta d\phi^2$). We represent $q_{AB}$ by a complex dyad (e.g. $q^A=(1,1/\sin\theta)$) and introduce the complex differential angular operators $\eth,\bar{\eth}$~\cite{Newman66}; see also \cite{Gomez97,Bishop2016a}. Then $h_{AB}$ is represented by the complex quantity $J=q^Aq^Bh_{AB}/2$ (with $J=0$ characterizing spherical symmetry), and we also introduce the complex quantity $U=U^Aq_A$.
Einstein's equations are
\begin{equation}
E_{ab}:=\;\; R_{ab}=8\pi(T_{ab}-\frac{1}{2}g_{ab}T) \,,
\end{equation}
with: $E_{ab}$ representing a set of {\bf equations}, rather than tensor components; $T_{ab}=\rho V_aV_b$ for dust of density $\rho$ and 4-velocity $V_a$; $T=-\rho$. Einstein's equations may be categorized as
\begin{align}
	\mbox{Hypersurface equations:  }&E_{11}\,,
	E_{1A}q^A\,,
	E_{AB}h^{AB}\,,\nonumber \\
	\mbox{Evolution equation:  }&E_J=E_{AB}(q^Aq^B-J h^{AB})\,,\nonumber \\
	\mbox{Constraint equations:  }&E_{00}\,,
	E_{01}\,,
	E_{0A}q^A\,.
\end{align}
Much previous work uses $E_J=E_{AB}q^Aq^b$; in the case that the background is not Minkowskian, the additional term introduced above leads to some simplification of the equation. A characteristic initial value problem may be formulated for the above with $J$ specified on a null hypersurface $u=$~constant~\cite{Isaacson83}.

\begin{figure}
	\begin{center}
		\includegraphics[width=0.50\columnwidth,angle=0]{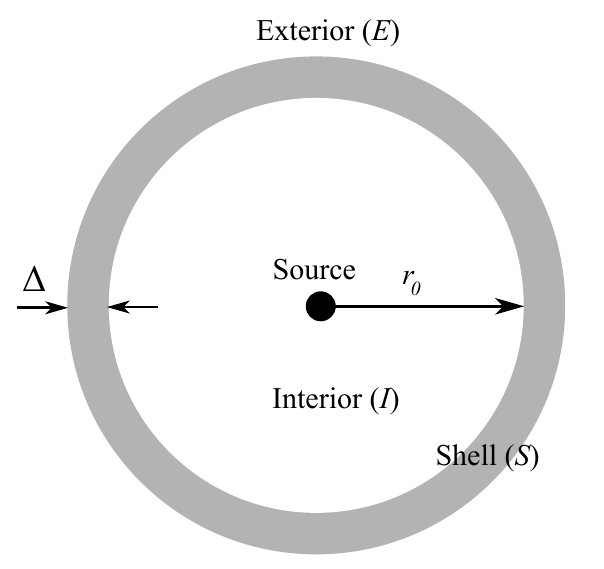}
		\caption{Schematic representation: The spacetime is empty apart from a GW source at the origin, and a shell of mass $M_S$ located between $r=r_0$ and $r=r_0+\Delta$. The spacetime thus comprises three regions as shown: Interior ($I$), Shell ($S$) and Exterior ($E$).
			\label{f-schematic}}
	\end{center}
\end{figure}


We consider the physical problem of a spacetime  that is empty except in a shell located at $r_0<r<r_0+\Delta$. where there is a spherically symmetric distribution of dust with a density profile that vanishes at $r=r_0$ and $r=r_0+\Delta$; see Fig.~\ref{f-schematic}. An example density profile is
\begin{equation}
\rho=\rho_c\left(\frac{1}{r^3}-\frac{r_0}{r^4}\right)
\left(\frac{r_0+\Delta}{r^4}-\frac{1}{r^3}\right)\,,
\label{e-ex_rho}
\end{equation}
and other density profiles tested are in Eq.~(\ref{e-other_rho}) in Appendix~\ref{a-compalg}. There is also a source of quadrupolar GWs at the origin, but the first problem that needs to be solved is the background solution for which this source is neglected. The problem is spherically symmetric so $J=U=0$. Further, the shell density is small and terms of ${\mathcal O}(\rho^2)$ are neglected. The collapse of the shell under its own gravity is an effect with acceleration ${\mathcal O}(\rho^2)$ and is therefore ignored. Thus the shell is treated as static, and the only non-zero metric coefficients are $\beta(r),W(r)$. The Einstein equations for $E_{11},E_{AB}h^{AB}$ then simplify to~\cite{Isaacson83,vanderwalt2010}
\begin{equation}
\partial_r\beta=2\pi r \rho (V_1)^2\,,\qquad
\partial_r W=e^{2\beta}-1-4\pi e^{2\beta}\rho r^2\,.
\end{equation}
Here, to ${\mathcal O}(\rho)$, $\rho (V^{[B]}_1)^2=\rho$ and $\beta={\mathcal O}(\rho)$, so we have
\begin{equation}
\partial_r\beta=2\pi r \rho \,,\qquad
\partial_r W=2\beta-4\pi \rho r^2\,,
\label{e-back}
\end{equation}
which are solved subject to the boundary conditions $\beta\rightarrow 0$ as $r\rightarrow \infty$ (so that the background coordinates are asymptotically Minkowskian), and $W={\mathcal O}(r)$ as $r\rightarrow 0$ (so that the $W/r$ is regular at the origin). We find:
\begin{align}
	r<r_0:&\;\; \beta^{[B]}=B_0\,,\qquad W^{[B]}=2r B_0 \nonumber \\
	r_0+\Delta<r:&\;\;\beta^{[B]}=0\,,\qquad W^{[B]}=-2M_S\,,
	\label{e-BWback}
\end{align}
with the superfix ${}^{[B]}$ indicating a quantity in the background depending only on $r$, and where $B_0$ is a constant and $M_S$ is the mass of the shell (see below). The solution for $\beta^{[B]},W^{[B]}$ in $r_0\le r\le r_0+\Delta$ is lengthy and is in the Supplementary Material, see Appendix~\ref{a-compalg}. In the case of Eq.~(\ref{e-ex_rho}) we find,
\begin{align}
	M_S=&\int^{r_0+\Delta}_{r_0}4\pi\rho r^2 dr=
	\rho_c\frac{\pi\Delta^3(10r_0^2+10r_0\Delta+3\Delta^2)}
	{15 r_0^4(r_0+\Delta)^4}\,,\nonumber\\
	B_0=&-\rho_c
	\frac{\pi\Delta^3 (2r_0+\Delta)(5r_0^2+5r_0\Delta+2\Delta^2)}
	{30 r_0^5(r_0+\Delta)^5}=-\frac{M_S}{2r_0}
	+{\mathcal O}(\Delta)\,,
	\label{e-MS-B0}
\end{align}
with the result $B_0=-M_S/2r_0+{\mathcal O}(\Delta)$ applying to all density profiles tested. The solution in $r>r_0+\Delta$ is Schwarzschild spacetime in Eddington-Finkelstein coordinates. The solution in $r<r_0$ is Minkowski spacetime (which in standard form has $\beta=W=0$), and this can be seen upon applying the coordinate transformation
\begin{equation}
u\rightarrow u^\prime =(1+2B_0)u\,.
\label{e-coord-t}
\end{equation}
We check, using computer algebra, that for a given density profile, $\beta$ and $W$ and their first derivatives are continuous at the interfaces at $r=r_0$ and $r=r_0+\Delta$; and also that the solution in the shell satisfies all 10 Einstein equations.

\section{Perturbed solution}
\label{s-pert}
Having determined the background solution, which can be regarded as Minkowski plus small spherically symmetric corrections of $\mathcal{O}(\rho)$, we perturb the solution by writing the metric quantities $(\beta,U,W,J)$ as
\begin{equation}
\beta=\beta^{[B]}+\beta^{[e]}\,,\;U=U^{[e]}\,,\;W=W^{[B]}+W^{[e]}\,,\;J=J^{[e]}\,,
\end{equation}
where the superfix ${}^{[e]}$ indicates the deviation  of a quantity about its background value.
The metric perturbations are functions of $(u,r,x^A)$ and are treated as ${\mathcal O}(\epsilon)$ with $\epsilon\ll 1$, and with the smallness parameters $\rho_c,\epsilon$ independent of each other. We also introduce the matter field perturbations
\begin{equation}
\rho=\rho^{[B]}+\rho^{[e]}\,,\;V_a=V_a^{[B]}+V_a^{[e]}\,,
\end{equation}
with $V_a^{[e]}$ treated as ${\mathcal O}(\epsilon)$ and $\rho^{[e]}$ treated as ${\mathcal O}(\rho_c\epsilon)$. Linearization about certain backgrounds was described in~\cite{Bishop-2005b}; using a similar procedure we retain only terms of order ${\mathcal O}(\rho_c,\epsilon,\rho_c\epsilon)$ and find that the hypersurface and evolution equations are
\begin{align}
E_{11} &: \,
\frac{4}{r}\,\left(\partial_r \beta^{[e]}
+  \,{ \partial_r \beta^{[B]}} \right)
= 8 \pi T_{11} \,,\nonumber \\
q^{a} E_{1A} &: \,\,
\frac{1}{2r^2} \, \left[ -2\,r^4 \partial_r \left(\frac{\eth \beta^{[e]}}{r^2} \right)
- 2 \, \partial_r\left(r^4\,\beta^{[B]} { \partial_r U^{[e]}} \right)	
+ \partial_{r} \left({r}^{4}{\partial_{r} U^{[e]}}\right)  
+ r^2 { \bar{\eth} \partial_r J^{[e]}}  \right] \nonumber \\
& = 8 \pi q^{A} T_{1A} \,,\nonumber \\
h^{AB}E_{AB}&:\,
4 \,{ \beta^{[B]}}
-2\,{ \eth \bar{\eth} \beta^{[e]}}	
-2\,{ \partial_r W^{[B]}}
+4 \,{ \beta^{[e]}} { \partial_r W^{[B]}} \nonumber \\
& - \left(2\,\beta^{[B]} - 1 \right) \left[ 4 \,{ \beta^{[e]}}
-2{ \partial_r W^{[e]}} 
+ \,\frac{1}{r^2} \, \partial_r \left({ r^{4} \eth  \bar{U}^{[e]}}
+ { r^{4} \bar{\eth} U^{[e]}} \right) \right]  \nonumber \\
& +\frac{1}{2}\, \left({ \bar{\eth}^2 J^{[e]}}
+ { \eth^2 \bar{J}^{[e]}} \right) \nonumber \\
& = 8 \pi \left( h^{AB}T_{AB} - r^2\,T \right)\,, \nonumber \\
E_J&:
-2\,{ \eth^2 \beta^{[e]} }
- \partial_r \left( { r \, \partial_r J^{[e]}}{ W^{[B]}} \right) \nonumber \\
& + \left(2\,{ \beta^{[B]}} -1 \right)\, \left[ -\partial_{r}\left({r}^2 { \eth  U^{[e]}} \right) 
+ \partial_{r}\left({r^2}{ \partial_{r} J^{[e]}} \right) 
- 2\,r\,\partial_r \left({r}{ \partial_u  J^{[e]}} \right)\right] \nonumber \\
& = 0\,.
\label{e-Einstein}
\end{align}

The above equations are tackled using the method of separation of variables, rather than formulation as a characteristic initial value problem. The required ansatz for the quantities $\beta^{[e]},U^{[e]},W^{[e]},J^{[e]},\rho^{[e]},V_a^{[e]}$ is
\begin{align}
	\beta^{[e]}=&\Re(\beta^{[2,2]}(r)e^{i\nu u}){}_0Z_{2,2}\,,\;\;
	U^{[e]}=\Re(U^{[2,2]}(r)e^{i\nu u}){}_1Z_{2,2}\,,\;\;
	W^{[e]}=\Re(W^{[2,2]}(r)e^{i\nu u}){}_0Z_{2,2}\,,\nonumber \\
	J^{[e]}=&\Re(J^{[2,2]}(r)e^{i\nu u}){}_2Z_{2,2}\,,\;\;
	\rho^{[e]}=\Re(\rho^{[2,2]}(r)e^{i\nu u}){}_0Z_{2,2}\,,\;\;
	V^{[e]}_0=\Re(V^{[2,2]}_0(r)e^{i\nu u}){}_0Z_{2,2}\,,\nonumber \\
	V^{[e]}_1=&\Re(V^{[2,2]}_1(r)e^{i\nu u}){}_0Z_{2,2}\,,\;\;
	q^AV^{[e]}_A=\Re(V^{[2,2]}_{ang}(r)e^{i\nu u}){}_1Z_{2,2}\,,
	\label{e-ansatz}
\end{align}
with the superfix ${}^{[2,2]}$ indicating a coefficient of ${}_s Z_{2,2}$.
The perturbations oscillate in time with frequency $\nu/(2\pi)$. The quantities ${}_s Z_{\ell,m}$ are spin-weighted spherical harmonic basis functions related to the usual ${}_s Y_{\ell,m}$ as specified in~\cite{Bishop-2005b,Bishop2016a}. They have the property that ${}_0 Z_{\ell,m}$ are real, enabling the description of the metric quantities $\beta,W$ (which are real) without mode-mixing; however, for $s\ne 0$ ${}_s Z_{2,2}$ is, in general, complex. A general solution may be constructed by summing over the $(\ell,m)$ modes, but that is not needed here, since we are considering a source that is continuously emitting GWs at constant frequency dominated by the $\ell=2$ (quadrupolar) components (Of course, the wave frequency changes with inspiral, and we are assuming that this timescale is much longer than the wave period). We substitute the ansatz Eqs.~(\ref{e-ansatz}) into Eqs.~(\ref{e-Einstein}) with $\ell=2$ to obtain: (1) On integrating over the sphere, only the $\ell=0$ part survives and Eqs.~(\ref{e-back}) for the background are obtained; (2) Multiplication by ${}_sZ_{2,m}^*$ (where ${}^*$ denotes complex conjugate) followed by integration over the sphere kills the spherically symmetric part, and Eqs.~(\ref{e-Einstein}) transform to a system of ordinary differential equations in $r$:
\begin{align}
	E_{11}:\,&\frac{4}{r}\, {\frac {d}{dr}}{ \beta^{[2,2]}}
	=8\,\pi \, \left( { \rho^{[2,2]}}-2\,{\rho^{[B]}}  { V_1^{[2,2]}} \right) \,,\nonumber \\
	q^AE_{1A}:&  \frac {1}{r^2}\left[ \,\sqrt {6}\, r^4 \, \frac {d}{dr} \left( \frac{\beta^{[2,2]}}{r^2} \right) \right.
	+ \,\frac{1}{r^{2}} \frac {d}{dr} \left( r^4 \, \beta^{[B]} \frac {d}{dr} U^{[2,2]} \right) \nonumber \\
	& - \frac{1}{2}\frac {d}{dr}\left({r}^{4}{ \frac {d}{dr} U^{[2,2]}}\right) 
	\left. + \,r^2 {\frac {d}{dr}}{ J^{[2,2]}} \right] \nonumber \\
	& =8\,\pi\,{ \rho^{[B]}} \, { V_{ang}^{[2,2]}}   \,,\nonumber \\
	h^{AB}E_{AB}:\,& 12\,{ \beta^{[2,2]}} 
	+ 4\,{ \beta^{[2,2]}}  {\frac {d}{dr}}{ W^{[B]}} 
	+ 2\,\sqrt {6}\,{ J^{[2,2]}} \nonumber \\
	& - \left(2\, \beta^{[B]} -1 \right)\left[4\,{ \beta^{[2,2]}}  
	- 2\,\sqrt {6}\,r\,{ U^{[2,2]}} 
	- \sqrt {6}\,\frac {d}{dr}\left(r^2\,{ U^{[2,2]}}\right)
	- 2\,\frac {d}{dr}{ W^{[2,2]}} \right] \nonumber \\
	& = 8\,\pi \,r^2\, \rho^{[2,2]} \,,\nonumber \\
	E_{J}: \,& \,\, 4\,\sqrt{6}{ \beta^{[2,2]}} 
	+\partial_r \left(r \, { W^{[B]}} \,  {\frac {d}{dr}}{ J^{[2,2]}} \right)   \nonumber \\
	& +\left(2\,\beta^{[B]} - 1 \right)\left[2\,{\frac {d}{dr}} \left({r}^{2} \,{ U^{[2,2]}} \right)  
	-{\frac {d}{dr}} \left({ {r}^{2} }  {\frac {d}{dr}}{ J^{[2,2]}}\right) 
	+2\,i\,{\nu}\,r\, {\frac {d}{dr}} \left( { r\, J^{[2,2]}} \right) \right] \nonumber \\
	& =0\,.
	\label{e-hyper-odes}
\end{align}
Eqs.~(\ref{e-hyper-odes}) contain terms of order ${\mathcal O}(\epsilon,\rho\epsilon)$; terms of ${\mathcal O}(\rho^2)$ may be larger than those of ${\mathcal O}(\rho\epsilon)$, but are excluded by the procedure that kills the spherically symmetric part of Eqs.~(\ref{e-Einstein}). In changing from Eqs.~(\ref{e-Einstein}) to Eqs.~(\ref{e-hyper-odes}), formulas were used for the effect of the $\eth$ operator on ${}_sZ_{\ell,m}$~\cite{Bishop-2005b,Bishop2016a}
\begin{align}
	&\eth\,{}_{-1}Z_{2,2}=-\sqrt{6} \, {}_{0}Z_{2,2}\,,\;\;
	\eth\,{}_{0}Z_{2,2}=\sqrt{6} \, {}_{1}Z_{2,2}\,,\;\;
	\bar{\eth}\,{}_{1}Z_{2,2}=-\sqrt{6} \, {}_{0}Z_{2,2}\,,\;\;
	\eth\,{}_{1}Z_{2,2}=2 \, {}_{2}Z_{2,2}\,,\;\;\nonumber \\
	& \bar{\eth}\,{}_{2}Z_{2,2}=-2 \, {}_{1}Z_{2,2}\,,\;\;
	\eth^2\,{}_{-1}Z_{2,2}=-6 \, {}_{1}Z_{2,2}\,,\;\;
	\eth^2\,{}_{0}Z_{2,2}=2 \sqrt{6} \, {}_{1}Z_{2,2}\,,\;\;
	\eth\bar{\eth}\,{}_0Z_{\ell,m}=-6\,{}_0Z_{\ell,m}\,,\;\; \nonumber \\
	&\eth\bar{\eth}\,{}_1Z_{\ell,m}=-6\,{}_1Z_{\ell,m}\,,\;\;
	\bar{\eth}^2\,{}_{2}Z_{2,2}=2\sqrt{6} \, {}_{0}Z_{2,2}\,,\;\;
	\eth^2\,{}_{-2}Z_{2,2}=2\sqrt{6} \, {}_{0}Z_{2,2}\,,
	\label{e-eth2Z}
\end{align}
where the above formulas are specialized to the case $\ell=2$. As well as the hypersurface Eqs.~(\ref{e-hyper-odes}), we will also need the constraints $E_{0a}$, which are
\begin{align}
	E_{00}:& -\frac{1}{2{r}^{3}}  \left[
	12\, r { \beta^{[2,2]}} 
	+24\, r { \beta^{[B]}}  { \beta^{[2,2]}}  
	-\frac{2}{r}{\frac {d}{dr}} \left({r}^{2}{\frac {d}{dr}}{ \beta^{[2,2]}}\right) \right. \nonumber \\
	& -{\sqrt {6}\,{r}\,{ U^{[2,2]}}}\left({\frac {d}{dr}} \left(\frac{ W^{[B]}}{r}\right)	
	+2\,{\frac {d}{dr}}{ \beta^{[B]}} \right) \nonumber \\
	& - 6\,{r^3} \,{ W^{[B]}} \,{\frac {d}{dr}} \left(\frac{ \beta^{[2,2]}}{r^2} \right) 
	-2\,{r}^{2}\left(2{\frac {d^{2}}{d{r}^{2}}}{ \beta^{[2,2]}} { W^{[B]}} 
	+{\frac {d}{dr}}{ \beta^{[2,2]}} {\frac {d}{dr}}{ W^{[B]}}\right) \nonumber \\
	&- 6\,{r^3} \,{ W^{[2,2]}} \,{\frac {d}{dr}}  \left(\frac{ \beta^{[B]}}{r^2}\right) 
	-2\,{r}^{2}\left(2 {\frac {d^{2}}{d{r}^{2}}}{ \beta^{[B]}} { W^{[2,2]}} 
	+  {\frac {d}{dr}}{ \beta^{[B]}} {\frac {d}{dr}}{ W^{[2,2]}} \right) \nonumber \\
	& +6\,{ W^{[2,2]}} 
	-{r}^{2}{\frac {d^{2}}{d{r}^{2}}}{ W^{[2,2]}}
	-r\,{ W^{[2,2]}}{\frac {d^{2}}{d{r}^{2}}}{ W^{[B]}}
	-\,r{ W^{[B]}}  {\frac {d^{2}}{d{r}^{2}}}{ W^{[2,2]}} 
	-2\,i{ \nu}\,r\,{ W^{[2,2]}}  \nonumber \\
	& \left. + i \,\frac{4\nu}{r^2}{\frac {d}{dr}}\left({r}\,{ \beta^{[2,2]}} \right)
	+4\,i\,{ \nu}\,r\,{ \beta^{[2,2]}}  { W^{[B]}} 
	+4\,i\,{ \nu}\,{r}^{2}\,{\frac {d}{dr}}{ \beta^{[2,2]}} { W^{[B]}} 	
	-2\,\sqrt {6}\,i\,{ \nu}\,{r}^{3}\,{ U^{[2,2]}}  \right] \nonumber \\
	& =4\,\pi \,\left( { \rho^{[2,2]}}+{ \rho^{[B]}} \frac { 2\beta^{[2,2]}r + W^{[2,2]}}{r} \right)\,,
	\nonumber \\
	E_{01}:& -\,\frac {1}{2{r}^{2}}  \left[24\,{ \beta^{[B]}} { \beta^{[2,2]}}
	+ 12\,{ \beta^{[2,2]}}  
	- r{\frac {d^{2}}{d{r}^{2}}}{ W^{[2,2]}} 
	+4\,i{ \nu}{r}^{2}\, {\frac {d}{dr}}{ \beta^{[2,2]}} \right. \nonumber \\
	& \left. - \frac {d}{dr} \left(2\,r^2 {\frac {d}{dr}}{ \beta^{[2,2]}} 
	+ 2\, r\,{ W^{[B]}}  {\frac {d}{dr}}{ \beta^{[2,2]}}
	+ 2\, r\,{ W^{[2,2]}} {\frac {d}{dr}}{ \beta^{[B]}} 
	+ r^2 \,\sqrt {6}{ U^{[2,2]}} \right) \right] \nonumber \\
	& =4 \,\pi \, \left( {\rho^{[2,2]}}-2 \,{ \rho^{[B]}}{ V_1^{[2,2]}}+{ \rho^{[B]}} \frac { W^{[2,2]}}{r} \right)\,,
	\nonumber \\
	q^AE_{0A}:\,&  \frac {1}{2{r}^{2}} \left[4\,{r}^{2}{ \beta^{[B]}}  { U^{[2,2]}}     
	+2\,{\frac {d}{dr}}\left({r}^{4}{ \beta^{[B]}}  {\frac {d}{dr}}{ U^{[2,2]}} \right) 	    
	-2\,i\,{ \nu}\,{r}^{4}{ \beta^{[B]}} \,{\frac {d}{dr}}{ U^{[2,2]}} \right. \nonumber \\
	& - \sqrt {6} \, \left( 2\, \,r\,{ W^{[2,2]}} {\frac {d}{dr}}{ \beta^{[B]}} 
	+\,r^2\,{\frac{d}{dr}} \left(\frac{W^{[2,2]}}{r} \right)
	-2\, \,i \,{ \nu}\,{r}^{2}\,{ \beta^{[2,2]}} \right) \nonumber \\
	& -2\,{r}^{2} {\frac {d}{dr}}{ W^{[B]}} { U^{[2,2]}}
	- \frac{1}{r}{\frac{d}{dr}}\left({r}^{4} { W^{[B]}} {\frac{d}{dr}}{ U^{[2,2]}} \right) \nonumber \\ 
	&\left.  -2\,{r}^{2}\,{ U^{[2,2]}} 
	-{\frac {d}{dr}}\left({r}^{4}\,{ U^{[2,2]}} \right) 	 
	+i\,{ \nu}\,{r}^{4}\,{\frac {d}{dr}}{ U^{[2,2]}} 	 
	+2\,i{ \nu}\,{r}^{2}\,{ J^{[2,2]}} \right] \nonumber \\
	& =4 \pi\, \rho^{[B]} \left( 2\,{ V_{ang}^{[2,2]}}+{ U^{[2,2]}}  {r}^{2} \right)\,.
	\label{e-constraint}
\end{align}

The procedure for solving Eqs~(\ref{e-hyper-odes}) and (\ref{e-constraint}) was given in~\cite{Bishop-2005b}, and in outline is as follows. Firstly, $E_{11}$ is integrated to find $\beta^{[2,2]}(r)$; then $q^AE_{1A}$ and $E_J$ are solved together to give $J^{[2,2]}(r),U^{[2,2]}(r)$; then $h^{AB}E_{AB}$ may be integrated to give $W^{[2,2]}(r)$. The solution obtained has $6$ constants of intergration, $2$ of which are fixed on applying the constraints $E_{00}, q^AE_{0A}$.

We will see later that the matter terms can be expressed as functions of $\rho^{[B]}$ and the metric, so that Eqs.~(\ref{e-hyper-odes}) take the form
\begin{equation}
(\mathcal{M}+\rho_c\mathcal{R})(f)=0\,,
\label{e-schematic}
\end{equation}
where $f$ is a multi-vector containing $\beta^{[2,2]}(r), J^{[2,2]}(r), U^{[2,2]}(r), W^{[2,2]}(r)$, and $\mathcal{M,R}$ are linear differential operators with $\mathcal{M}$ being the operator when $\rho_c=0$, i.e. when the background is Minkowski. Let $f_M$ be the solution to the homogeneous problem $\mathcal{M}(f_M)=0$. $f_M$ was derived in Eq.~(56) of~\cite{Bishop-2005b} (but note that the reference used a different notation, and used $\eth^s\,{}_0Z_{\ell,m}$ rather than ${}_sZ_{\ell,m}$ as angular basis), and
is
\begin{align}
	\beta^{[2,2]}=&b_0\,, \nonumber \\
    W^{[2,2]}=&6i\nu r^2 C_{10}+r(12C_{10}-10b_0)+C_{50}
	-\frac{12i\nu C_{40}}{r}-\frac{6C_{40}}{r^2}
	-C_{in0}\exp(2ir\nu)\frac{3}{r^2}\,,\nonumber \\
	U^{[2,2]}=&-\sqrt{6}i\nu C_{10}+\frac{2\sqrt{6} b_0}{r}
	+\frac{2\sqrt{6} C_{30}}{r^2}-\frac{4i\nu\sqrt{6} C_{40}}{r^4}
	-\frac{3\sqrt{6} C_{40}}{r^4} \nonumber \\
	&-C_{in0}\exp(2ir\nu)\sqrt{6}\left(i\frac{\nu}{r^3}
	-\frac{3}{2r^4}\right)\,,\nonumber \\
	J^{[2,2]}=&2\sqrt{6}C_{10}+\frac{2\sqrt{6}C_{30}}{r}
	+\frac{2\sqrt{6}C_{40}}{r^3}+C_{in0}\exp(2ir\nu)\sqrt{6}\left(
	\frac{1}{r^3}-2i\frac{\nu}{r^2}-\frac{\nu^2}{r}\right)\,.
	\label{e-pert}
\end{align}
The solution includes constants of integration $b_0, C_{in0}, C_{10}, C_{30}, C_{40}, C_{50}$, two of which are fixed on applying the constraints $E_{00},q^AE_{0A}$ giving
\begin{equation}
C_{50}=12\nu^2 C_{40}\,,\;\;
C_{10}=\frac{2b_0+i\nu C_{30}+i\nu^3C_{40}}{3}\,.
\label{e-cons}
\end{equation}
The remaining Einstein equation $E_{01}$ is known as the trivial equation, since it is automatically satisfied provided all the other Einstein equations are satisfied~\cite{Bondi62}. The solution Eq.~(\ref{e-pert}) subject to Eq.~(\ref{e-cons}) will be denoted by $f_M$; and this may be specialized to the case of no incoming radiation by setting $C_{in0}=0$ with the solution denoted by $f_{M0}$.

The gravitational news ${\mathcal N}$ is defined in a coordinate system that satisfies the Bondi gauge conditions $\lim_{r\rightarrow\infty}J,U,\beta,W/r =0$, and is calculated on making the required coordinate transformation. The procedure in the general case was described in~\cite{Bishop97b}, which was then simplified for the linearized approximation in~\cite{Bishop-2005b} (Sec.3.3); an explicit expression for the news was given in~\cite{Reisswig:2006}, Eq.~(16). Allowing for the conventions used here, we find that for $f_{M0}$, ${\mathcal N}_{M0}=-\sqrt{6}\nu^3 \Re(iC_{40}\exp(i\nu u))\,{}_2Z_{2,2}$. The gravitational wave strain and news are related (see~Eq.~(276) in \cite{Bishop2016a}) $\mathcal{H}_{M0}=h_+ +ih_\times = 2\int{\mathcal N}_{M0} du$, giving
\begin{equation}
{\mathcal H}_{M0}=\Re(H_{M0} \exp(i\nu u))\,{}_2Z_{2,2}\;\mbox{with}\;\; H_{M0}=-2\sqrt{6}\nu^2 C_{40}\,.
\label{e-HM0}
\end{equation}
$C_{40}$ is determined by the physical problem being modelled, and $b_0,C_{30}$ represent gauge freedoms; e.g., for an equal mass $m$ binary with orbital radius $r_o$, $C_{40}=2mr_o^2\sqrt{\pi/15}$~\cite{Bishop:2011}.

We regard the solution $f_{M0}$ as applying when no matter shell is present, i.e. when $\rho_c=0$, with all the constants of order ${\mathcal O}(\epsilon)$ and with $C_{40}$ fixed by the physics of the source. Then the solution in the presence of the matter shell will be $f_{M0}$ plus small corrections of order ${\mathcal O}(\epsilon\rho_c)$. We construct a solution in each of the three regions $r<r_0,r_0<r<r_0+\Delta,r_0+\Delta<r$, and then apply matching conditions at the boundaries $r=r_0,r=r_0+\Delta$.

\subsection{Solution inside the shell, i.e. $r<r_0$}
\label{s-I}
Perturbations in this region are on a background that is explicitly Minkowskian in $(u^\prime,r,\theta,\phi)$ coordinates; thus the solution $f_M$ applies, although the effect of boundary conditions may be to change the values of the constants by ${\mathcal O}(\epsilon\rho_c)$ from their values in $f_{M0}$. We therefore make the substitutions in Eq.~(\ref{e-pert}) $b_0\rightarrow b_0+b_{0I},C_{in0}\rightarrow C_{inI},C_{10}\rightarrow C_{10}+C_{1I},C_{30}\rightarrow C_{30}+C_{3I},C_{40}\rightarrow C_{40}+C_{4I},C_{50}\rightarrow C_{50}+C_{5I}$ with $b_{0I}$ etc. of order ${\mathcal O}(\epsilon\rho_c)$, and denote the solution as $f_{I^\prime}$. The solution $f_I$ in global $(u,r,\theta,\phi)$ coordinates is obtained on applying the coordinate transformation Eq.~(\ref{e-coord-t}) to $f_{I^\prime}$; using computer algebra, we have evaluated $f_I$ explicitly  and give the formulas in the Supplementary Material, see Appendix~\ref{a-compalg}. Using computer algebra, the solution $f_I$ has been substituted into Eqs.~(\ref{e-hyper-odes}) with $\beta^{[B]},W^{[B]}$ taking the values for $r<r_0$ in Eqs.~(\ref{e-BWback}) and (\ref{e-MS-B0}), and we have confirmed that the equations are satisfied. Values of the constants $C_{1I},C_{5I}$ have been determined by substituting $f_I$ into Eqs.~(\ref{e-constraint}), and then the constraints have been re-evaluated to confirm that they are satisfied.

\subsection{Solution within the shell, i.e. $r_0<r<r_0+\Delta$}
The metric perturbations introduce perturbations into the matter fields, and these act as source terms in the Einstein equations. We evaluate the condition that the energy momentum tensor should be divergence-free, $\nabla_c T_{ab} g^{bc}=0$, with the metric and metric connection terms evaluated using $f_{M0}$ (rather than $f$). The result is four equations for five unknowns $\rho^{[2,2]}(r),V^{[2,2]}_0(r),$ $V^{[2,2]}_1(r),V^{[2,2]}_{ang}(r)$ (note that $V^{[2,2]}_{ang}(r)$ is complex with two components), and the system is closed on applying the condition that the 4-velocity has unit norm, i.e. $V_aV_bg^{ab}=1$. The resulting formulas are given in the Supplementary Material, see Appendix \ref{a-compalg}, and we note that they are explicit functions of $r$ depending on the parameters $b_0,C_{10},C_{30},C_{40},C_{50},r_0,\Delta,\rho_c$. We also consider terms on the left hand side of Eqs.~(\ref{e-hyper-odes}) of the form ($\beta^{[B]}$ or $W^{[B]}$)$\times$(metric perturbation) and use $f_{M0}$ rather than $f$ to evaluate the metric perturbation; thus these terms also become explicit functions of $r$ depending on known parameters.

The justification for using $f_{M0}$ instead of $f$ is that the error introduced is ${\mathcal O}(\rho_c\epsilon)$, and all such terms are multiplied by a term of ${\mathcal O}(\rho_c)$ so the total error is ${\mathcal O}(\rho_c^2\epsilon)$ which is ignorable. The result of these simplifications is that Eq.~(\ref{e-schematic}) becomes
\begin{equation}
\mathcal{M}f=-\rho_c\mathcal{R}_S(f_{M0})\,,
\end{equation}
for which the solution to the homogeneous part has already been obtained, and there is just a particular integral that needs to be found;  $\mathcal{R}_S$ denotes the specific form of the operator $\mathcal{R}$ within the shell. We denote the solution within the shell as $f_S$, and introduce additional constants $b_{0S},C_{inS},C_{1S},C_{3S},C_{4S},C_{5S}$ similarly to what was done in Sec.~\ref{s-I} with $f_I$. The explicit solution for $f_S$ is obtained by means of computer algebra; it is very long and is given in the Supplementary Material, see Appendix~\ref{a-compalg}. Using
computer algebra, we have evaluated the constraints to find expressions for $C_{1S},C_{5S}$, and the solution $f_S$ has been confirmed by checking that it satisfies Eqs.~(\ref{e-hyper-odes}) and (\ref{e-constraint}) to order ${\mathcal O}(\rho_c\epsilon)$.

\subsection{Solution exterior to the shell, i.e. $r_0+\Delta<r$}

In this region, the background geometry is Schwarzschild and perturbations about this background are not known analytically, but a numerical solution has been obtained~\cite{Bishop2009}. Here, however, we can use the same procedure as in the previous sub-section, that is for terms of the form ($\beta^{[B]}$ or $W^{[B]}$)$\times$(metric perturbation) we use $f_{M0}$ rather than $f$ to evaluate the metric perturbation, and Eq.~(\ref{e-schematic}) becomes $\mathcal{M}f=-\rho_c\mathcal{R}_E(f_{M0})$.  We denote the solution outside the shell as $f_E$, and introduce additional constants $b_{0E},C_{inE},C_{1E},C_{3E},C_{4E},C_{5E}$ similarly to what was done in Sec.~\ref{s-I} with $f_I$. The explicit solution for $f_E$ is obtained by means of computer algebra and is given in the Supplementary Material, see Appendix~\ref{a-compalg}. Using
computer algebra, we have evaluated the constraints to find expressions for $C_{1E},C_{5E}$, and the solution $f_E$ has been confirmed by checking that it satisfies Eqs.~(\ref{e-hyper-odes}) and (\ref{e-constraint}) to order ${\mathcal O}(\rho_c\epsilon)$.

Having determined the solution in the exterior region, we can now evaluate the gravitational news ${\mathcal N}^{[2,2]}$, and the formula is given in the Supplementary Material, see Appendix~\ref{a-compalg}. It involves $C_{40},C_{4E},\rho_c,r_0,\Delta$ and we need to express $C_{4E}$ in terms of the physical variables $C_{40},\rho_c,r_0,\Delta$ so as to complete the calculation.

\subsection{Matching conditions and the complete solution}

After having used the constraints in each region, $f_E,f_S,f_I$ contain 12 free constants. The condition of no incoming radiation in the exterior requires that the coefficient of $\exp(2ir\nu)/r$ in $f_E$ be zero, leading to
\begin{equation}
C_{inE}=-4\nu \pi C_{40} M_S\,,
\end{equation}
and continuity of the metric quantities $\beta,W,J,U$ at $r=r_0$ and $r=r_0+\Delta$ imposes 8 conditions
\begin{align}
	\beta^{[2,2,I]}(r=r_0)&=\beta^{[2,2,S]}(r=r_0)\,,\qquad
	\beta^{[2,2,S]}(r=r_0+\Delta)=\beta^{[2,2,E]}(r=r_0+\Delta)\,,
	\nonumber\\
	J^{[2,2,I]}(r=r_0)&=J^{[2,2,S]}(r=r_0)\,,\qquad
	J^{[2,2,S]}(r=r_0+\Delta)=J^{[2,2,E]}(r=r_0+\Delta)\,,\nonumber\\
	U^{[2,2,I]}(r=r_0)&=U^{[2,2,S]}(r=r_0)\,,\qquad
	U^{[2,2,S]}(r=r_0+\Delta)=U^{[2,2,E]}(r=r_0+\Delta)\,,
	\nonumber\\
	W^{[2,2,I]}(r=r_0)&=W^{[2,2,S]}(r=r_0)\,,\qquad
	W^{[2,2,S]}(r=r_0+\Delta)=W^{[2,2,E]}(r=r_0+\Delta)\,,
	\label{e-boundary}
\end{align}
where the $I,S,E$ within the superfix indicates a quantity in the interior, shell or exterior respectively. We first solve the continuity conditions at $r=r_0+\Delta$ to find expressions for $b_{0E},C_{3E},C_{4E},C_{inS}$, and then we solve the continuity conditions at $r=r_0$ to find expressions for $b_{0S},C_{3S},C_{4S},C_{inI}$; all these expressions are given in the Supplementary Material, see Appendix~\ref{a-compalg}.
Since the Einstein hypersurface equations contain second derivatives of $J$ and $U$, we also expect $\partial_r J,\;\partial_r U$ to be continuous at $r=r_0$ and $r=r_0+\Delta$. Using computer algebra we have checked that, provided Eqs.~(\ref{e-boundary}) are satisfied, the following conditions are also satisfied
\begin{align}
	\partial_rJ^{[2,2,I]}(r=r_0)&=\partial_rJ^{[2,2,S]}(r=r_0)\,,\qquad
	\partial_rJ^{[2,2,S]}(r=r_0+\Delta)=\partial_rJ^{[2,2,E]}(r=r_0+\Delta)\,,\nonumber\\
	\partial_rU^{[2,2,I]}(r=r_0)&=\partial_rU^{[2,2,S]}(r=r_0)\,,\qquad
	\partial_rU^{[2,2,S]}(r=r_0+\Delta)=\partial_rU^{[2,2,E]}(r=r_0+\Delta)\,.
\end{align}

The outcome is that three degrees of freedom remain, $b_{0I},C_{3I},C_{4I}$, and the expression for ${\mathcal H}$ involves $C_{40},C_{4I}$. As already discussed, the value of the constant $C_{40}$ is determined by the physics of the GW source, and we need another physical condition in order to close the system and to fix $C_{4I}$. The calculation of the perturbations from an equal mass binary assumes no incoming radiation, but the back-reaction due to the matter distribution may introduce such a term. Thus, the computer algebra script for the calculation of GWs emitted by an equal mass binary~\cite{Bishop:2011} has been amended with $C_{in0}\ne 0$  but ${\mathcal O}(\rho_c)$. We find
\begin{equation}
C_{4I}=-\frac{C_{inI}}{2}\,.
\label{e-C4I-Cin0}
\end{equation}
This result applies independently of source properties, and in particular it applies when the mass of the orbiting binary is zero. Thus, it is reasonable to regard the result in Eq.~(\ref{e-C4I-Cin0}) as general, representing the reflection at the origin of an incoming GW.

Then, finally, for all density profiles tested, the gravitational wave strain in terms of the mass of the shell $M_S$ and as measured by an observer at future null infinity, was found to be
\begin{equation}
{\mathcal H}=
\Re\left(H_{M0}\left(
1+\frac{2M_S}{r_0}+\frac{2iM_S}{r_0^2\nu}+\frac{i M_Se^{-2ir_0\nu}}{2r_0^2\nu}+{\mathcal O}\left(\frac{M_S\Delta}{r^2_0},\frac{M_S}{r_0^3\nu^2} \right)\right)\exp(i\nu u)\right)\,{}_2Z_{2,2}\,,
\label{e-Nshell}
\end{equation}
where $H_{M0}$ was given in Eq.~(\ref{e-HM0}). Each of the terms containing $M_S$ in Eq.~(\ref{e-Nshell}) represents a correction to the wave strain in the absence of the shell, and these correction terms are discussed in Sec.~\ref{s-phys} below.

The formula Eq.~(\ref{e-Nshell}) was derived using quite complicated computer algebra scripts, and it is important to investigate its reliability through consistency checks:
\begin{itemize}
	\item In each of the three regions, we confirmed using computer algebra that all 10 Einstein equations are satisfied. The four constraint equations were not used in the construction of the solution, although they were used to fix two (in each region) constants. Note that it was found that if an error were introduced into one of the hypersurface equations, then the formulas obtained for $C_1,C_5$ were not constant but were functions of $r$. Thus, the fact that the solutions satisfy the constraints amounts to a strong consistency test.
	\item The solution was constructed to ensure continuity of $\beta,W,J,U$ at $r=r_0$ and $r=r_0+\Delta$, but continuity of $\partial_r J, \;\partial_r U$ is also needed.  We confirmed using computer algebra that these conditions are satisfied, which again amounts to a strong consistency test.
	\item As will be discussed in the next section, the formula Eq.~(\ref{e-Nshell}) satisfies physical expectations, with one of the terms in the formula being derivable in an independent way.
\end{itemize}

\section{Physical interpretation}
\label{s-phys}

We investigate the physical meaning of each of the correction terms in Eq.~(\ref{e-Nshell}), in the light of the expectation that the dust shell cannot add or remove energy from the GWs:
\begin{itemize}
	\item Correction 1, $2M_S/r_0$. In natural Minkowski coordinates $u^\prime$ and in the absence of a matter shell, $H^\prime=-2\nu^{\prime 2}\sqrt{6}C_{40}$. From Eq.~(\ref{e-coord-t}), it follows that $\nu^\prime=\nu (1-2B_0))$, and thus from Eq.~(\ref{e-MS-B0}), $\nu^\prime=\nu (1+M_S/r_0))$. Thus $H^\prime\approx -2\nu^2\sqrt{6}C_{40}(1+2M_S/r_0)$. This term is therefore a consequence of the coordinate transformation Eq.~(\ref{e-MS-B0}), i.e. it represents the gravitational red-shift effect of the shell.
	\item Correction 2, $2iM_S/(r_0^2\nu)$. The term is out of phase with the leading terms $1+2M_S/r_0$. Thus, to ${\mathcal O}(M_S)$, the magnitude of ${\mathcal H}$, and therefore of the energy of the GW, is not affected; but the shell does change the phase of the GW.
	\item Correction 3, $iM_S e^{-2ir_0\nu}/(r_0^2\nu)$. This term does affect the magnitude of ${\mathcal H}$. We have checked using computer algebra (see Appendix \ref{a-compalg}) that if ${\mathcal H}$ is calculated with $C_{inI}$ set to zero the term disappears, and so it is interpreted as being an effect due to the shell generating an incoming GW. Such an incoming GW modifies the geometry near the source  and affects the radiation reaction (or self-force) and thus the inspiral rate. The calculation of the self-force is a $2^{\mbox{\small nd}}$ order effect which is beyond the scope of this work. The energy change in the GW at infinity can therefore be interpreted as being caused by the modification of the self-force, rather than by energy being transferred to or from the dust cloud.
\end{itemize}

In order for Correction 1 to be a measurable, $M_S/r_0$ would need to have a significant value (but note that the results obtained here also assume that $(M_S/r_0)^2$ is negligible). For Corrections 2 and 3 to be measurable, we would also need that $r_0\nu$ should not be large, where $r_0\nu=2\pi r_0/\lambda$ with $r_0/\lambda$ being the number of wavelengths in $r_0$; thus the shell would need to be very close to the GW source.

We now consider the case of a thick shell comprising a dust cloud of constant density ($\rho_0$) extending from near the origin to an observer at $r=r_F$, so that $\delta M_S=4\pi r_0^2 \rho_0 \delta r_0$; integration gives
\begin{equation}
{\mathcal H}=
\Re\left(H_{M0}\left(
1+4\pi \rho_0 r_F^2+4 i\rho_0 r_F\lambda+{\mathcal O}\left(\frac{\rho_0\log(r_F)}{\nu^2} \right)\right)\exp(i\nu u)\right)\,{}_2Z_{2,2}\,,
\label{e-Nshellthick}
\end{equation}
and the integral of Correction 3 is omitted since it is smaller than the order term. Applying Eq.~(\ref{e-Nshellthick}) to cosmology, and taking the cosmological density as $10^{-29}$g/cm${}^3\approx 0.7 \times10^{-8}$Mpc${}^{-2}$ in geometric units, it is found that, apart from the gravitational redshift, no measurable effect is expected since the GW wavelength $\lambda$ is at most ${\mathcal O}(10^4)$km $\ll 1$Mpc.

\section{Conclusion}
\label{s-conc}

Using the Bondi-Sachs formalism, this work has investigated within linearized perturbation theory, the effect of a spherical dust shell on  GWs sourced from the center of the shell. It was found that the GWs were modified, although without any energy transfer between the GWs and the shell. This finding is novel. In the context of cosmology, the effects are too small to be measurable; but the effect would be measurable if a GW event were to occur with a source surrounded by a massive shell and with the radius of the shell and the wavelength of the GWs of the same order.

There are three avenues for further work. Firstly, the matter equation of state needs to be generalized beyond dust to include shear viscosity, and perhaps other forms of dissipation. Secondly, the background spacetime without the matter shell can be changed from Minkowski to something more appropriate to cosmology. Solutions are known for de Sitter spacetime~\cite{Bishop2016}; and for Einstein-de Sitter~\cite{vanderwalt2010}, although in this case the algebraic complexity of the solution may make the construction of perturbative solutions problematic. Finally, we found that a shell of matter, in effect, reflects part of an outgoing GW, and so in the context of a burst (rather than a continuous) source it is possible that an echo would be the result. Now, it should be possible to model a burst source as a Fourier sum of the eigensolutions obtained here, i.e. as something of the form $\sum_n a_n f(\nu_n)$, and in this way to investigate GW echoes.


\appendix

\section{Description of computer algebra scripts}

\label{a-compalg}
The computer algebra scripts used in this paper are written in Maple in plain text format, and are available
as Supplementary Material.
Note that the output files may be viewed using a plain text editor provided line-wrapping is switched off.

The scripts \texttt{gamma.out, initialize.map, lin.map, ProcsRules.map} are not used directly, but are called by the other scripts described below. \texttt{gamma.out} contains formulas for the Bondi-Sachs metric, its inverse and the metric connection coefficients.  \texttt{lin.map} constructs the Einstein equations linearized about a given background. \texttt{ProcsRules.map} contains various procedures and rules that are used by other scripts.

The script \texttt{initialize.map} initializes various arrays etc., and sets the density profile of the matter as given in Eq.~(\ref{e-ex_rho}); other density profiles tested were
\begin{align}
	\rho&=\rho_c\left(\frac{1}{r^3}-\frac{r_0}{r^4}\right)
	\left(\frac{(r_0+\Delta)^2}{r^5}-\frac{1}{r^3}\right)\,,\qquad
	\rho=\rho_c\left(\frac{1}{r^4}-\frac{r_0}{r^5}\right)
	\left(\frac{r_0+\Delta}{r^4}-\frac{1}{r^3}\right)\,,\nonumber\\
	\rho&=\rho_c\left(\frac{1}{r^2}-\frac{r_0}{r^3}\right)
	\left(\frac{(r_0+\Delta)}{r^5}-\frac{1}{r^4}\right)\,,\qquad
	\rho=\rho_c\left(\frac{1}{r^2}-\frac{r_0}{r^3}\right)
	\left(\frac{r_0+\Delta}{r^6}-\frac{1}{r^5}\right)\,.
	\label{e-other_rho}
\end{align}

The script \texttt{backgroundShell.map} constructs the background (spherically symmetric) solution for the given density profile; it also checks that the metric functions are sufficiently smooth at the interface $r=r_0$, and that the solution satisfies all $10$ Einstein equations. The output is in  \texttt{backgroundShell.out}

The script \texttt{paperEqs.map}, with output in \texttt{paperEqs.out}, generates the formulas given in Eqs.~(\ref{e-Einstein}), (\ref{e-hyper-odes}) and (\ref{e-constraint}); note that some manual simplifications have been applied to the formulas generated by the computer algebra.

The script \texttt{shell.map} uses the divergence-free condition on the energy-momentum tensor, $\nabla_a T_{bc}g^{ac}=0$, to determine the fluid properties, i.e. density and velocity perturbations. It then constructs the metric in $r<r_0,r_0<r<r_0+\Delta$ and $r>r_0$, and checks that the solutions obtained satisfy all $10$ Einstein equations. Finally, it constructs and solves the continuity conditions at $r=r_0,r=r_0+\Delta$, and then evaluates the gravitational wave strain. The output is in \texttt{shell.out}.

The script \texttt{shellCinI\_0.map} is the same as \texttt{shell.map} except that: on line 399 \texttt{CinI} is hard-coded to be 0, and some output has been suppressed. The output is in \texttt{shellCinI\_0.out}.

The script \texttt{regular\_0\_IncomingGW.map} is an adaptation of the script \texttt{regular\_0.map} used in~\cite{Bishop:2011} to calculate the GWs emitted by an equal mass binary. The adaptations are: (a) an incoming wave, as a free parameter, is included; and (b) the coefficient names have been changed to be consistent with those used in \texttt{shell.map}. The output is in \texttt{regular\_0\_IncomingGW.out}.

The file \texttt{formulas.pdf} is in pdf format and contains the Maple output for
\begin{align*}
	&\beta^{[B]},W^{[B]},
	\beta^{[2,2,I]},J^{[2,2,I]},U^{[2,2,I]},W^{[2,2,I]},C_{1I},C_{5I},
	\rho^{[2,2]},V^{[2,2]}_0,V^{[2,2]}_1,V^{[2,2]}_{ang}
	\nonumber\\ &\beta^{[2,2,S]},J^{[2,2,S]},U^{[2,2,S]},W^{[2,2,S]},C_{1S},C_{5S},
	\beta^{[2,2,E]},J^{[2,2,E]},U^{[2,2,E]},W^{[2,2,E]},C_{1E},C_{5E},
	\nonumber\\
	&{\mathcal N}^{[2,2]},b_{0E},C_{3E},C_{4E},C_{inS},
	b_{0S},C_{3S},C_{4S},C_{inI}\,.
\end{align*}
The formulas for the above are generated during the execution of \texttt{shell.map} and written to the file \texttt{formulas.out}.

The file \texttt{Eqs.pdf} is in pdf format and includes the content of \texttt{paperEqs.out} with annotations, together with the formulas for $\beta^{[B]},W^{[B]}$ within the matter shell extracted from \texttt{backgroundShell.out}.

\begin{acknowledgements}
This work was supported by the National Research Foundation, South Africa, under grant numbers 118519 and 114815.
\end{acknowledgements}

%
 \section*{Conflict of interest}

 The authors declare that they have no conflict of interest.

\bibliographystyle{spphys}       

\bibliography{aeireferences,t}

\begin{thebibliography}{10}
\providecommand{\url}[1]{{#1}}
\providecommand{\urlprefix}{URL }
\expandafter\ifx\csname urlstyle\endcsname\relax
  \providecommand{\doi}[1]{DOI \discretionary{}{}{}#1}\else
  \providecommand{\doi}{DOI \discretionary{}{}{}\begingroup
  \urlstyle{rm}\Url}\fi

\bibitem{Esposito71b}
F.P. {Esposito}, The Astrophysical Journal \textbf{168}, 495 (1971).
\newblock \doi{10.1086/151103}

\bibitem{Ehlers87}
J.~Ehlers, A.R. Prasanna, R.A. Breuer, Classical and Quantum Gravity
  \textbf{4}(2), 253 (1987).
\newblock \doi{10.1088/0264-9381/4/2/009}.
\newblock \urlprefix\url{https://doi.org/10.1088%2F0264-9381%2F4%2F2%2F009}

\bibitem{Ehlers96}
J.~Ehlers, A.R. Prasanna, Classical and Quantum Gravity \textbf{13}(8), 2231
  (1996).
\newblock \doi{10.1088/0264-9381/13/8/016}.
\newblock \urlprefix\url{https://doi.org/10.1088%2F0264-9381%2F13%2F8%2F016}

\bibitem{Hawking66}
S.W. {Hawking}, The Astrophysical Journal \textbf{145}, 544 (1966).
\newblock \doi{10.1086/148793}

\bibitem{Esposito71a}
F.P. {Esposito}, The Astrophysical Journal \textbf{165}, 165 (1971).
\newblock \doi{10.1086/150884}

\bibitem{Madore73}
J.~Madore, Communications in Mathematical Physics \textbf{30}(4), 335 (1973).
\newblock \doi{10.1007/BF01645508}.
\newblock \urlprefix\url{https://doi.org/10.1007/BF01645508}

\bibitem{Anile78}
A.M. Anile, V.~Pirronello, Il Nuovo Cimento B (1971-1996) \textbf{48}(1), 90
  (1978).
\newblock \doi{10.1007/BF02748651}.
\newblock \urlprefix\url{https://doi.org/10.1007/BF02748651}

\bibitem{Prasanna99}
A.~Prasanna, Physics Letters A \textbf{257}(3), 120  (1999).
\newblock \doi{https://doi.org/10.1016/S0375-9601(99)00313-8}.
\newblock
  \urlprefix\url{http://www.sciencedirect.com/science/article/pii/S0375960199003138}

\bibitem{Goswami17}
G.~Goswami, G.K. Chakravarty, S.~Mohanty, A.R. Prasanna, Phys. Rev. D
  \textbf{95}, 103509 (2017).
\newblock \doi{10.1103/PhysRevD.95.103509}.
\newblock \urlprefix\url{https://link.aps.org/doi/10.1103/PhysRevD.95.103509}

\bibitem{Baym17}
G.~Baym, S.P. Patil, C.J. Pethick, Phys. Rev. D \textbf{96}, 084033 (2017).
\newblock \doi{10.1103/PhysRevD.96.084033}.
\newblock \urlprefix\url{https://link.aps.org/doi/10.1103/PhysRevD.96.084033}

\bibitem{Bondi62}
H.~Bondi, M.G.J. van~der Burg, A.W.K. Metzner, Proc. R. Soc. London
  \textbf{A269}, 21 (1962)

\bibitem{Sachs62}
R.~Sachs, Proc. Roy. Soc. London \textbf{A270}, 103 (1962)

\bibitem{Winicour05}
J.~{Winicour}, Living Reviews in Relativity \textbf{8}, 10 (2005)

\bibitem{Bishop2016a}
N.T. Bishop, L.~Rezzolla, Living Rev. Relativ. \textbf{19}, 1 (2016).
\newblock \doi{DOI 10.1007/s41114-016-0001-9}.
\newblock \urlprefix\url{http://dx.doi.org/10.1007/s41114-016-0001-9}

\bibitem{Winicour83}
J.~{Winicour}, Journal of Mathematical Physics \textbf{24}, 1193 (1983)

\bibitem{Bishop96}
N.T. Bishop, R.~G{\'o}mez, L.~Lehner, J.~Winicour, Phys. Rev. D \textbf{54},
  6153 (1996)

\bibitem{Isaacson85}
R.~Isaacson, S.~Welling, J, J.~Winicour, Journal of Mathematical Physics
  \textbf{26}(5), 2859 (1985).
\newblock \doi{10.1063/1.526712}

\bibitem{Bishop-2005b}
N.T. Bishop, Class. Quantum Grav. \textbf{22}(12), 2393 (2005).
\newblock \doi{10.1088/0264-9381/22/12/006}

\bibitem{Bishop2016}
N.T. Bishop, Phys. Rev. D \textbf{93}, 044025 (2016).
\newblock \doi{10.1103/PhysRevD.93.044025}.
\newblock \urlprefix\url{https://link.aps.org/doi/10.1103/PhysRevD.93.044025}

\bibitem{Ashtekar-2015-2}
A.~Ashtekar, B.~Bonga, A.~Kesavan, Phys. Rev. D \textbf{92}, 044011 (2015)

\bibitem{Ashtekar-2015-3}
A.~Ashtekar, B.~Bonga, A.~Kesavan, Phys. Rev. D \textbf{92}, 10432 (2015)

\bibitem{Conklin2017}
R.S. Conklin, B.~Holdom, J.~Ren, Phys. Rev. D \textbf{98}, 044021 (2018).
\newblock \doi{10.1103/PhysRevD.98.044021}.
\newblock \urlprefix\url{https://link.aps.org/doi/10.1103/PhysRevD.98.044021}

\bibitem{Konoplya2019}
R.A. Konoplya, Z.~Stuchl\'{\i}k, A.~Zhidenko, Phys. Rev. D \textbf{99}, 024007
  (2019).
\newblock \doi{10.1103/PhysRevD.99.024007}.
\newblock \urlprefix\url{https://link.aps.org/doi/10.1103/PhysRevD.99.024007}

\bibitem{Newman66}
E.T. Newman, R.~Penrose, Journal of Mathematical Physics \textbf{7}(5), 863
  (1966).
\newblock \doi{10.1063/1.1931221}.
\newblock \urlprefix\url{https://doi.org/10.1063/1.1931221}

\bibitem{Gomez97}
R.~G{\'o}mez, L.~Lehner, P.~Papadopoulos, J.~Winicour, Class. Quantum Grav.
  \textbf{14}(4), 977 (1997)

\bibitem{Isaacson83}
R.~Isaacson, J.~Welling, J.~Winicour, J. Math. Phys. \textbf{24}, 1824 (1983)

\bibitem{vanderwalt2010}
P.J. van~der Walt, N.T. Bishop, Phys. Rev. D \textbf{82}, 084001 (2010).
\newblock \doi{10.1103/PhysRevD.82.084001}.
\newblock \urlprefix\url{https://link.aps.org/doi/10.1103/PhysRevD.82.084001}

\bibitem{Bishop97b}
N.T. {Bishop}, R.~{G{\'o}mez}, L.~{Lehner}, M.~{Maharaj}, J.~{Winicour}, Phys.
  Rev. D \textbf{56}, 6298 (1997).
\newblock \doi{10.1103/PhysRevD.56.6298}

\bibitem{Reisswig:2006}
C.~{Reisswig}, N.T. {Bishop}, C.W. {Lai}, J.~{Thornburg}, B.~{Szilagyi},
  Classical and Quantum Gravity \textbf{24}, 327 (2007).
\newblock \doi{10.1088/0264-9381/24/12/S21}

\bibitem{Bishop:2011}
N.T. Bishop, D.~Pollney, C.~Reisswig, Class. Quantum Grav. \textbf{28}, 155019
  (2011)

\bibitem{Bishop2009}
N.T. Bishop, A.S. Kubeka, Phys. Rev. D \textbf{80}, 064011 (2009).
\newblock \doi{10.1103/PhysRevD.80.064011}.
\newblock \urlprefix\url{https://link.aps.org/doi/10.1103/PhysRevD.80.064011}

\end{thebibliography}

\end{document}